# Not at Home on the Range: Peer Production and the Urban/Rural Divide

**Isaac L. Johnson\*, Yilun Lin\*, Toby Jia-Jun Li†\*, Andrew Hall\*, Aaron Halfaker§,
Johannes Schöning‡, Brent Hecht\***

\*GroupLens Research, Department of Computer Science, University of Minnesota, †Human-Computer Interaction Institute, Carnegie Mellon University, §Wikimedia Foundation, ‡Expertise Center for Digital Media, Hasselt University - tUL - iMinds

{ijohnson, ylin, hall, bhecht}@cs.umn.edu, tobyli@cs.cmu.edu, ahalfaker@wikimedia.org, johannes.schoening@uhasselt.be

**ABSTRACT**
Wikipedia articles about places, OpenStreetMap features, and other forms of peer-produced content have become critical sources of geographic knowledge for humans and intelligent technologies. In this paper, we explore the effectiveness of the peer production model across the rural/urban divide, a divide that has been shown to be an important factor in many online social systems. We find that in both Wikipedia and OpenStreetMap, peer-produced content about rural areas is of systematically lower quality, is less likely to have been produced by contributors who focus on the local area, and is more likely to have been generated by automated software agents (i.e. "bots"). We then codify the systemic challenges inherent to characterizing rural phenomena through peer production and discuss potential solutions.

**ACM Classification Keywords**
H.5.m. Information interfaces and presentation (e.g., HCI): Miscellaneous;

## INTRODUCTION
Peer-produced content has been a game-changer in the tremendously important domain of geographic information. We now learn about places near and far by reading peer-produced Wikipedia articles, and many of the maps we use on a daily basis leverage peer-produced data from OpenStreetMap (OSM), the "Wikipedia of maps" [12,33,67,74]. Behind the scenes, many important intelligent algorithms utilize data from Wikipedia and OSM to make geographic inferences about the world [20,27,44].

The importance of geographically-referenced peer-produced content, also known as peer production volunteered geographic information or *peer production VGI* [31,57], has led some researchers to inquire as to whether the peer production content generation model works equally well in describing all types of geographies (e.g., [19,28,43,52]). However, missing from this literature is a robust analysis of the effectiveness and character of peer production across the urban-to-rural spectrum. Researchers in HCI have shown that urban/rural dynamics can play prominent roles in a variety of online social systems ranging from social networks [17] to photo-sharing sites [32] to check-in platforms [32]. These results echo decades of research in the social sciences that chronicles differences in how urban and rural areas have adopted and used technology (e.g., [7,37]).

The goal of this paper is to address this gap in the literature by examining peer production's relative ability to generate quality content about urban and rural areas. Because our goal is to understand urban/rural dynamics in peer production generally rather than in a single type of peer production community, we consider both Wikipedia and OpenStreetMap. These are two of the largest and most impactful peer production communities and are communities with substantially different approaches to peer production. For the same reason, we also examine content about countries with two very different human geographies: the United States and China, which are often the focus of cross-cultural analyses (e.g., [70]). Studying urban/rural content generation in diverse online and offline communities allows us to gain a richer understanding of the phenomena of interest [3].

We find that *regardless of the peer production community or country, content about rural areas is of substantially lower quality than urban areas*. For instance, Wikipedia articles about rural places are much more likely to be assessed as low quality by Wikipedia contributors than articles about urban places, rural OpenStreetMap entities (e.g., buildings, roads, etc.) have fewer tags than urban entities, and Wikipedia articles about rural areas have substantially more content written by editors who have not specialized in the local area. Indeed, our results show that Wikipedia articles about very rural places in the U.S. have less than 5% of their content (on average) written by specialized editors, whereas this number is over 37% in urban areas.

Our results also highlight the important role of "bots" (i.e. automated software agents) and automation-assisted batch editors in creating content about rural areas. For instance, whereas 4.5% (median) of content in Wikipedia articles about urban places is bot- or batch editor-generated, the corresponding proportion for rural places is over 23%. Although these automated solutions generate low-quality content, we show how they are critical to peer production's ability to provide a baseline level of coverage in rural areas.

Rural areas have significantly different sociodemographic characteristics than urban areas [8], and the models we use in this research control for these potential confounds. Through these modeling efforts, this research also tangentially uncovers new findings about systemic peer production biases in our control variables. For instance, we find that English Wikipedia articles about places with a higher percentage of Democratic votes are of higher quality, and the same is true of more educated places in the Chinese Wikipedia.

As we will detail below, these results have highly tangible implications for (1) human consumers of peer-produced content (e.g., Wikipedia readers, users of OSM-based maps) and (2) the many systems and algorithms that rely on peer-produced content to understand the world's geography. With regard to humans, our results mean that articles about rural areas and OSM-based maps about rural areas are of substantially lower quality (on average) than those about urban places. With regard to systems and algorithms, our results suggest that, to many peer production-based geographic technologies, rural areas frequently "look the same" and are defined only by their topology and government census data.

This research is also the first to characterize and quantify the fundamental challenge facing peer production communities' efforts to map and describe rural phenomena: in rural areas, the ratio of entities of interest (e.g., incorporated towns) to potential local editors can be orders of magnitude higher than in urban areas. Since many contributors of geographic peer-produced information contribute about places local to them (e.g., [26,29]) and local contributions are generally of higher quality (e.g., [8], our coding study below), rural areas are systemically disadvantaged in the peer production content generation model. We show how this disadvantage manifests itself in the results we identify in this paper. Fortunately, our results also point to potential partial solutions to this issue, and we expand on these implications for design below.

To summarize, this paper makes the following contributions:

- We establish that peer-produced content about urban areas is of higher quality than that about rural areas, demonstrating that this difference persists across two prominent peer production communities (Wikipedia and OpenStreetMap) with very different communication and collaboration structures and two countries with very different human geographies.
- We highlight the critical role that bots and batch editing tools play in ensuring that there is any content at all about some rural areas.
- We discuss how our results reveal systemic challenges facing peer production projects in describing and mapping rural phenomena and highlight how our results can inform the design of potential sociotechnical solutions.
- Finally, through controlling for sociodemographic factors, we identify new content biases in peer-production related to politics, education, and profession.

## RELATED WORK

This paper draws motivation from work in three areas: (1) coverage biases in peer production, (2) urban/rural dynamics in online social systems, and (3) content localness in peer production. Below, we discuss each of these areas in more detail, with other related work discussed in context in later sections.

### Coverage Biases in Peer Production

This research is informed by a thread of related work that looks at systemic variations in peer-produced content along dimensions other than the urban/rural spectrum. A major recent thrust of this work relates to gender dynamics, with several studies showing that both OSM and the English Wikipedia have more content about and for men than about and for women (e.g., [32,38,45,55,63]). Language has been shown to be a particularly strong factor in Wikipedia coverage biases, with each language edition having much better coverage of places where the corresponding language is spoken (e.g., [18,27,28,57]). Other factors behind systemic variations in content include national culture [49,51,53], and politics [21].

### Urban/Rural Dynamics in Online Social Systems

The vast majority of work that considers issues associated with online social systems and population density has focused exclusively on urban areas [17,32]. The small body of online social systems research that has looked at differences between urban and rural areas has largely focused on social media, which has significantly different collaboration and contribution characteristics than peer production. For instance, Gilbert et al. found that rural MySpace users had significantly fewer friends than urban users [17]. Similarly, Hecht and Stephens identified that rural areas have fewer tweets, check-ins, and geotagged Flickr photos per capita than urban areas [32].

In work that provided key motivation for the research in this paper, several researchers have observed that

OpenStreetMap data seems to display different characteristics in rural and urban areas. For instance, Zielstra et al. [73] identified that OSM coverage was more extensive closer to a selection of German cities, and a similar finding was identified in the London area by Mashhadi et al. [43]. One goal of this paper is to build on these findings, most importantly by examining OSM and Wikipedia in a consistent, robust analytical framework so as to gain an understanding of peer production in rural areas as a whole, but also by focusing specifically on urban/rural dynamics and doing so across entire countries. This allows us to incorporate important control variables, use more sophisticated urban/rural sociodemographic statistics, and introduce other targeted approaches and metrics, ultimately resulting in the series of contributions with important implications outlined above.

**Localness and Peer Production**

Important context for this work also comes from the small literature on the localness of peer-produced content. This literature has shown that, in general, edits to geographic Wikipedia articles tend to come from people who are located (and likely live) near the subject of the article. For instance, Hecht and Gergle [29] found that over 25% of edits to Wikipedia articles about places in the English Wikipedia comes from within 100km and a similar finding was identified by Hardy et al. [26]. Related research on OpenStreetMap has shown that local editors tend to do higher quality work (something we confirm in our studies below). For example, Zielstra et al. [72] found that OSM editors contributed a higher diversity of edits in their home region and Eckle [8] found that familiarity with an area led to more accurate OSM mapping in a controlled experiment.

Recent work by Sen et al. [57] examined the geographic variability in these overall localness results, finding that Wikipedia articles about places in poorer countries and countries with a less healthy publishing industry have fewer local editors and reference fewer local sources. A portion of our research extends the work of Sen et al. to a more granular scale, looking at whether localness also varies across the urban/rural spectrum and finding that rural places have less local content, even on a per capita basis.

**DATA AND METRICS**

This research involves a number of different datasets and a variety of metrics, with many of these datasets and metrics being the output of somewhat complex processes. We first describe basic information about our Wikipedia and OSM datasets. Next, we discuss each of our other datasets and metrics in detail, grouped by whether they help us quantify (1) rural/urban dynamics, (2) peer-production quality, (3) peer-production quantity, or (4) control sociodemographic variables.

**Peer Production Datasets**

Most of our Wikipedia data was extracted from the static XML dumps of the English and Chinese language editions of Wikipedia using the WikiBrain API [56]. We analyzed the English Wikipedia when considering the United States and the Chinese Wikipedia when considering China. Like nearly all prior geographic Wikipedia work (e.g., [18,19,28,29,39]), we focus on "geotagged articles", or articles that have been tagged with a latitude and longitude location by Wikipedia editors. In total, we identified 218,709 English geotagged articles about places in the contiguous United States and 46,124 Chinese geotagged articles about places in China. Note that because of the requirements of our geographic modeling techniques (see next section), we only consider the contiguous 48 states when examining the United States.

Our OpenStreetMap data comes from the static Planet.osm dump from February 2014 for the United States and July 2015 for China. For the contiguous US, the dump contains 494 million nodes, 32.8 million ways, and 263 million tags. In China, there are 22.6 million nodes, 1.6 million ways, and 5.5 million tags.

We aggregate all Wikipedia and OpenStreetMap metrics to the county level in the contiguous United States (3,109 counties) and the prefecture level in China (344 prefectures). The census data necessary to perform our analyses at a more local scale in China is not available from the Chinese government (see below). As is considered best practice [30], we filter out geotagged Wikipedia articles about entities with footprints larger than a county/prefecture (e.g., articles about states, countries, continents) to avoid assigning their content to the single county/prefecture that contains the lat/lon of their geotag. It is important to point out that this aggregation allows us to make claims about *all data about places in a country/prefecture*, rather than the specific Wikipedia article about a county/prefecture or the specific OSM geometry describing the county/prefecture. This aggregative approach has often been shown to be effective in related work (e.g., [19,27,28,43,51]).

**Urban/Rural Datasets**

We obtained data about geographic variation in urbanness from government sources. In the United States, we make use of a statistic from the 2010 U.S. Census that describes the percentage of the population in a given county that lives in an urban area (*% Pop Urban*). This percentage is calculated using the U.S. Census' definition of urban areas, which includes both significant cities as well as urbanizations of 2,500 or more people [62]. When our analyses require discrete classifications along the urban/rural spectrum, we utilize the National Center for Health Statistics' (NCHS) urban-rural classifications [34], which assign each U.S. county an ordinal code from "1" (core urban) to "6" (entirely rural). New York County is assigned a "1", for example, and Loving County, Texas (the lowest-population county in the United States) is assigned a "6". In China, our *% Pop Urban* statistic comes from the Chinese government's 2010 Population Census. Urban areas are defined based on administrative districts and mainly include highly commercialized and populous districts.

**Content Quality Data and Metrics**

There are many definitions of content quality in peer-produced datasets, with each definition serving an important constituency and providing a unique view into the effectiveness of the content. In order to gain as deep an understanding as possible of quality variation across the urban/rural divide, we sought to examine both repositories using a variety of quality definitions aimed at uncovering different dimensions of peer production quality. Many quality metrics are country- or repository-specific due to the requirements of the methods by which they are calculated.

*WikiProject Quality Assessments*

In the English Wikipedia, most articles are assessed by members of the Wikipedia community with a quality score from an ordinal seven-point scale that ranges from "stub" class ("provides very little meaningful content") to "featured article" class ("a definitive source for encyclopedic information") [75]. These assessments have been used in a number of research projects as a holistic measure of the multifaceted notion that is Wikipedia article quality (e.g., [36,61,64–66]). We evaluate the quality differences between urban and rural areas using this rich quality metric by measuring the percentage of articles about places in a U.S. county that are assessed at C-class or higher (*% C-class or higher*). We use C-class as the threshold as Wikipedia's documentation describes it as the lowest quality class in which articles are still "useful to a casual reader" [75]. While the Chinese Wikipedia does have an analogous quality scale, it has not been validated in the literature to our knowledge, and, as such, we restrict this metric to the English Wikipedia/United States.

*Tag Richness*

In addition to describing the geometries of geospatial entities, OpenStreetMap also contains a large dataset of tags corresponding to these entities. OpenStreetMap tags are how humans and computers understand the semantics of the underlying geometries. Without them, maps (and algorithms) based on OSM data would not be able to, for instance, distinguish a hospital from a bar or a highway from a dirt road [60]. Tags also support OSM-based location-based services by providing them with venue opening hours information, cuisine type, and many other attributes. In general, the more tags an entity has, the more useful it is to humans and computers. To operationalize this notion of quality, we use a metric called "*Tag Richness*", which is defined simply as the average number of tags per entity in a county/prefecture.

*Content Diversity*

One quality metric that plays an important role in the value both humans and computers gain from peer-produced content is the amount of unique information available about a specific place. For instance, boilerplate Wikipedia articles about a town that merely describe the town's neighboring towns and basic census statistics are less useful to readers and algorithms/systems than articles that have rich descriptions of the town's unique history and character.

While the value of diverse content to readers is obvious, the value for systems/algorithms is more complex (but equally important). Systems/algorithms that use peer-produced knowledge typically use data models derived from article/region content (e.g., "bag of links" models for Wikipedia [1,46,47]), leveraging these data models to answer queries, assess the similarity of concepts, and support many other applications (e.g., [10,11,13,14,69]). If articles all have roughly the same content, the power of these systems/algorithms to discriminate between different places will be adversely affected, likely reducing application effectiveness in rural areas in some cases.

To operationalize content diversity, we use a metric we call *Outlink Entropy*, which has the advantage of being directly linked to the diversity of commonly-used "bag of links" models as well as capturing a human-visible notion of content diversity. A straightforward application of information entropy [58], outlink entropy measures the extent to which the links on pages about places in a county/prefecture all point to the same small set of articles, or whether they point to a diverse group of articles. For instance, if a large proportion of links in a county's geotagged articles point to the "United States Census" article (because a large proportion of the articles' content amounts to basic census statistics), this would result in low outlink entropy. In a higher entropy county, articles' links would point to a diverse set of other articles relevant to the county's history, current events, and so on.

*Ratio of Human-generated Content*

As has been described in prior work by Geiger and others (e.g., [15,16,25,71]), peer production communities are often complex ecosystems that consist of human editors and automated and partially-automated software agents. These agents "promote consistency in the content, structure, and presentation" of articles [61], and, in some cases, generate content. Much of this automated content generation usually amounts to the importing of pre-existing data or statistics into the genre and format of the peer production community. For example, many geotagged Wikipedia articles have text that a bot generated from census statistics, e.g., in the "Clayton, Missouri" article, there is bot-produced text that reads "As of the census of 2010, there were 15,939 people, 5,322 households, and 2,921 families residing in the city." Similarly, in OSM, large quantities data have been imported from the U.S. government's TIGER/Line street dataset.

The content generated by automated and partially-automated agents is often considered by members of peer production communities to be of substantially lower quality than that generated by humans. Indeed, in both OSM and Wikipedia, extensive debates have taken place as to whether automated content generation should continue, and, if so, whether and how it should be constrained [40,71]. To capture this notion of quality, we measured the amount of content contributed in each county/prefecture by human editors versus that contributed by automated or partially-automated agents.

Following prior work on identifying bots in Wikipedia [65], we distinguished human Wikipedia editors from bot editors by comparing the editor's username to usernames in the "bot" user group as well as by searching the username for the word "bot". Batch editors are identified by doing a case-insensitive search for the names of two very common editors, "AWB" and "WPCleaner", an approach that has been used successfully in prior work [35].

In OpenStreetMap, a feature was identified as bulk uploaded if its most recent edit was from a changeset in which edits occurred at a rate of faster than one per second and at a volume greater than 1000 edits. This approach is similar to prior work [71,72]. We report values from this classification but also tested our data with more relaxed criteria, which produced very similar results. Notably, using this metric, features that were initially uploaded in bulk but have since been edited in a sufficiently small or slow changeset are classified as human-edited.

Measuring the amount of content attributable to a specific editor or class of editors (e.g., bots vs. humans) in Wikipedia is non-trivial. Most past work has used the number or ratio of edits, but edits can be of different sizes, can be of malicious intent (e.g., vandalism), and so on. This is a known issue in the literature, and, to address it, we turn to the work of Halfaker et al. [24] in tracking the persistence of words through revisions of Wikipedia pages. We process the entire edit history for each geotagged article and compute the percentage of tokens (i.e. words, numbers) in the final version of the page that were contributed by each type of contributor (bots, batch editors, and humans).

*Content Contributed by "Local Experts"*
An important recent thread of Wikipedia research has adopted as a quality metric the extent to which the content in geotagged articles is coming from local experts (e.g., [19,57]). This research is motivated by recent studies (e.g., [8,72]) and by prominent geographer Michael Goodchild's claim in his formative article on VGI that "the most important value of volunteered geographic information may lie in what it can tell us about local activities" [13].

All prior localness work in the Wikipedia domain has relied heavily on IP geolocation, and nearly all of this work has examined localness at the country-to-country scale (i.e. a contributor is "local" if she is from the same country as the entity she is editing/contributing). Because the accuracy of IP geolocation declines tremendously when attempting to position IP addresses to their state, county, city, etc. rather than their country [50], the approaches for quantifying local expertise in prior work cannot be used for our more granular analyses (analyses that have been called for by some of this prior work, e.g., [57]).

To address this problem, we instead assess the percentage of Wikipedia article tokens about a given U.S. county that come from contributors who have exhibited some degree of local focus on that county. This has the benefit of allowing an editor to be considered a "local expert" in more than one county (e.g., their home county and the county in which they attend university), while at the same time filtering out edits by "fly-by" editors (we compare "local focus" and "fly-by" edits below). Specifically, as our definition of local expertise, we measure the percentage of tokens about a county that come from editors who have focused 10% or more of their effort on that county (as measured by edits to geotagged articles). This more multifaceted definition of localness is similar to the "$n$-days" metric that has been used when studying urban/rural dynamics in social media VGI [32,41]. However, because this definition is not directly comparable with past definitions of local expertise, we do not describe this quality metric as a "local expertise" metric, but rather a metric that measures the degree of "local spatial focus".

To confirm that editors who display local focus contribute different types of information than fly-by editors and to better understand the nature of each group's contributions more generally, we performed a small qualitative coding exercise. Two coders examined all tokens contributed by editors with local spatial focus and fly-by editors on 25 randomly selected articles about places in NCHS = "6" counties. The coders classified these tokens into four categories: (1) bot-like structured data (i.e. tokens that describe data from large, well-known external repositories like the U.S. Census), (2) structured data from local sources (i.e. tokens that describe data from a very local government agency), (3) administrative edits (e.g., typo fixes, syntax fixes), and (4) rich local information (e.g., information about the area's history, culture, or well-known persons).

The coders overlapped on tokens for 5 articles and achieved a relatively good Cohen's $\kappa$ = 0.69. As expected, the proportion of tokens that contained rich local information was much higher for editors with a local focus (61.3%) than for fly-by editors (24.0%) ($z$ = 5.65, $p$ < 0.001). Moreover, over 39% of fly-by editors' tokens were bot-like in nature compared to just 16% in those with a local focus ($z$ = 3.39, $p$ < 0.001).

**Content Quantity**
In addition to assessing variations in quality across the rural/urban spectrum, we also examine variations in the raw amount of content across this spectrum. We do so for several reasons. First and foremost, by examining the number of entities in urban and rural areas, we can assess the per capita "burden" in rural areas relative to that in urban areas. Secondly, the vast majority of existing work on biases in peer production repositories – especially those in the Wikipedia domain – has looked exclusively at quantity metrics (e.g., number of edits per capita in sub-Saharan African countries vs. Europe [19], length of articles about women vs. length of articles about men [55]). As such, analyzing the variation of these metrics in rural and urban areas has two additional benefits: (1) it affords comparability with this existing work and (2) as we will see, our results will reveal flaws in using raw content quantity alone as a metric in peer production bias research.

We selected our specific content quantity metrics by identifying metrics that are commonly used in the peer production bias literature (e.g., [4,19,27,38]). In Wikipedia, we look at number of articles per capita, number of outlinks per capita, and article bytes (length) per capita. In OSM, we examine nodes (points) per capita, ways (lines and polygons) per capita, and total tags per capita.

### Sociodemographic Control Variables

In both the U.S. and China, the human geography of rural and urban areas has sociodemographic differences that go well beyond population density. For instance, in the U.S., rural areas tend to be older, poorer, and vote more Republican [6,37]. In China, rural areas are poorer, less educated, and have a higher proportion of males [59]. In this research, we adopt two parallel perspectives on these associations. The first attempts to control for these factors, teasing out a purer effect for rural and urban (using *multivariate* models). The second considers rural areas as they are today (e.g., on average poorer, older, more Republican), incorporating their entire human geography in our assessment of variation in peer production content across the urban and rural spectrum (using *univariate* models).

The specific sociodemographic controls we consider in the U.S. are household median income (*HMI*), median age (*Median Age*), the percent of the population that is White and non-Latino (*% WNL*, a commonly used statistic in race and ethnicity work), the 2012 vote rate for Obama (*% Democratic*), and the percent of the population employed in management, business, science, or the arts (*% White Collar*). These data come from the 2010 U.S. Census (*% WNL*, *Median Age*), the 2009-2013 U.S. Census American Community (*HMI*, *% White Collar*), and *The Guardian* (*% Democratic*). The controls we included for China are gender ratio (*% Male*), the percent of the population that is not of Han ethnicity (*% Non-Han*), the percent of the population that is 15-64 years old (*% Age 15-64*), and the percent of the population that is college-educated (*% College or More*). All these data come from the 2010 Sixth National Population Census.

Other potential confounds (e.g., education in the U.S.) were not possible to include because of excessive collinearity with existing variables that would have destabilized the model coefficients or caused excessive positive skew, as with the very high broadband penetration rates in the United States. We also note that we included a dummy variable reflecting the presence of land managed by the National Park Service initially as a control (e.g., national parks). We anticipated that this information would help to distinguish between rural areas of two significantly different functions and characters. However, we found including this control only minimally changed the effects of the other variables and therefore removed it from the analysis framework.

### METHODS

Once the metrics described above had been calculated or collected, the remainder of our methodological approach consisted of a relatively straightforward univariate and multivariate regression-based modeling exercise (with the exception that our models need to account for spatial autocorrelation; see below). Our peer production quality and quantity metrics are our dependent variables and *% Pop Urban* and the other sociodemographic variables are our independent variables. We ran a separate regression for each dependent variable.

We log-transform variables as necessary to achieve normality. We then z-score scale all variables so that the resulting beta coefficients as produced by the regressions are directly related to unit standard deviation changes in the dependent variable. This approach allows for comparison of relative effect sizes between different variables. We test for spatial autocorrelation and run spatial regressions using the *spdep* package in R [5] according to spatial statistics best practices, which call for selecting one of two spatial regression models (error or lag) with fit test statistics [2].

### RESULTS

Table 1 contains the results of our spatial autoregressive models. Each row in Table 1 corresponds to one of the quality and quantity metrics defined above, and the cells of the table are populated with normalized effect sizes for the independent variables.

Table 1 tells a striking high-level story: examining the *% Pop Urban* columns (columns 4 and 5), we see that nearly across the board, *content in peer production repositories about urban areas is significantly different than content about rural areas*. With the exception of the multivariate results for China (a point to which return later), *% Pop Urban* is significant for almost all attributes (quantity and quality) in both repositories and both countries, and its normalized effect size is often very high. Overall, it appears that peer-produced information about urban and rural areas is of substantially different character.

Below this high-level story there are a number of critical themes that emerge from Table 1. The remainder of this section is dedicated to highlighting these themes.

### Theme 1: Urban Advantage in Quality

Table 1 reveals a strong and pervasive *pro-urban* bias when it comes to our quality metrics (for which the "Type" column = "Quality"). Whether we define quality by manually-assessed Wikipedia quality ratings, content diversity metrics, tag richness, local focus, or human production, *urban peer-produced content appears to be of significantly and substantially higher quality than rural content*. Aside from a few outliers, this result holds for both the univariate and multivariate regressions, across both countries, and across both repositories.

Unpacking the normalized effect sizes into their absolute values, the strength of the urban quality advantage becomes clearer. For instance, for every standard deviation (31.4% absolute) increase in *% Pop Urban*, our univariate regressions indicate that there is a 47.6% relative increase in the percentage of articles in that county that are assessed as C-class or better (41.2% when controlling for sociodemographics through our multivariate regressions).

When considering the percentage of content that comes from editors with a local spatial focus on a given U.S. county (as defined above), we see equally large effects. In purely rural counties (NCHS classification = "6"), *only 4%* of all tokens on Wikipedia pages come from these focused editors, who, as we saw above, contribute nuanced, local information at a much higher rate than "fly-by" editors (and of course bots and batch edits). The equivalent figure for core urban counties (NCHS classification = "1") is over nine times higher at 37.6%. Along the same lines, whereas 4.5% (median) of the tokens in articles about core urban counties are contributed by bots or batch editors, the equivalent number for entirely rural counties (NCHS code = "6") is 23.4%. In many of these counties, bots and batch editors generated over 60% of their content.

Table 1 also reveals a strong urban bias when it comes to content diversity, which is an important metric for both human and machine consumers of peer-produced content. It appears that rural counties have less unique content and more boilerplate information (e.g., census data), limiting the ability of people and machines to determine the unique character of places in these counties.

Interestingly, in many cases the urban bias in quality even persists when controlling for population. For instance, there are far more C-class or better articles *per capita* in urban areas. In other words, there are far more articles that are "useful [at least] to a casual reader" on a per capita basis in urban areas than in rural areas. Our univariate results indicate that with every standard deviation increase in *% Pop Urban* (31.6%), there is a corresponding 24.2% increase in the number of C-class or better articles per capita. We see a similar effect for local tokens: locally-focused editors are contributing fewer tokens on a per-capita basis in rural areas than their urban counterparts, which is likely one cause of the per-capita deficiency of C-class articles. These results point to a systemic underrepresentation of rural editors in Wikipedia, a point to which we return in the discussion section.

| UNITED STATES | | | | | | | | | |
|---|---|---|---|---|---|---|---|---|---|
| DATASET / METRIC | | | UNIVARIATE | MULTIVARIATE | | | | | |
| Comm-unity | Attribute | Type | % Pop Urban | % Pop Urban | HMI | Median Age | % Democratic | % White, Non-Latino | % "White-Collar" |
| Wikipedia | Outlink Entropy | Quality | 0.34*** | 0.29*** | -0.07*** | -0.01 | -0.10*** | -0.12*** | 0.10*** |
| Wikipedia | % C-Class Articles (or better) | Quality | 0.31*** | 0.27*** | 0.00 | 0.00 | 0.15*** | 0.03 | 0.13*** |
| Wikipedia | C-Class Articles (or better) per capita | Quality | 0.13*** | 0.12*** | -0.03 | 0.03 | 0.20*** | 0.10** | 0.09*** |
| Wikipedia | % Local Focus Tokens | Quality | 0.29*** | 0.24*** | -0.04* | -0.02 | 0.10*** | 0.06*** | 0.15*** |
| Wikipedia | Local Focus Tokens per Capita | Quality | 0.11*** | 0.09*** | 0.03 | 0.04* | 0.13*** | 0.09*** | 0.06*** |
| Wikipedia | % Human Tokens | Quality | 0.31*** | 0.26*** | -0.02 | -0.01 | 0.08** | -0.03 | 0.12*** |
| Wikipedia | Human Tokens per Capita | Quality | -0.42*** | -0.37*** | -0.17*** | 0.17*** | 0.20*** | 0.08** | 0.15*** |
| OSM | Tags per Feature | Quality | 0.20*** | 0.18*** | -0.07*** | -0.05** | -0.01 | 0.05** | -0.01 |
| OSM | % Human Nodes | Quality | 0.32*** | 0.26*** | 0.10*** | -0.05* | 0.15*** | 0.12*** | 0.13*** |
| OSM | % Human Ways | Quality | 0.30*** | 0.21*** | 0.09*** | -0.09*** | 0.17*** | 0.10** | 0.05** |
| Wikipedia | Articles per Capita | Quantity | -0.51*** | -0.42*** | -0.16*** | 0.19*** | 0.14*** | 0.06* | 0.07*** |
| Wikipedia | Length (Bytes) per Capita | Quantity | -0.47*** | -0.41*** | -0.18*** | 0.17*** | 0.19*** | 0.09*** | 0.13*** |
| Wikipedia | Outlinks per Capita | Quantity | -0.45*** | -0.37*** | -0.14*** | 0.19*** | 0.17*** | 0.07** | 0.10*** |
| OSM | Nodes per Capita | Quantity | -0.50*** | -0.41*** | -0.04* | 0.15*** | -0.06** | -0.05* | 0.01 |
| OSM | Ways per Capita | Quantity | -0.51*** | -0.41*** | -0.07*** | 0.18*** | -0.04 | -0.05* | 0.03* |
| OSM | Tags (Nodes + Ways) per Capita | Quantity | -0.54*** | -0.43*** | -0.08*** | 0.19*** | -0.05* | -0.05* | 0.01 |
| PEOPLE'S REPUBLIC OF CHINA | | | | | | | | | |
| DATASET / METRIC | | | UNIVARIATE | MULTIVARIATE | | | | | |
| Comm-unity | Attribute | Type | % Pop Urban | % Pop Urban | % Male | % Non-Han | % Age 15-64 | % College or More | |
| Wikipedia | Outlink Entropy | Quality | 0.28*** | 0.00 | 0.04 | 0.25** | -0.08 | 0.38*** | |
| OSM | Tags per Feature | Quality | 0.20*** | 0.10 | -0.08 | -0.13 | 0.14 | 0.04 | |
| OSM | % Human Nodes | Quality | 0.22*** | -0.13 | -0.10 | 0.05 | 0.22** | 0.09 | |
| OSM | % Human Ways | Quality | 0.12* | 0.10 | -0.06 | 0.13* | 0.15 | -0.06 | |
| Wikipedia | Articles per Capita | Quantity | -0.19*** | -0.34*** | 0.16*** | 0.30*** | -0.17* | 0.19** | |
| Wikipedia | Length (Bytes) per Capita | Quantity | -0.14*** | -0.11 | 0.18*** | 0.33*** | -0.18* | 0.36*** | |
| Wikipedia | Outlinks per Capita | Quantity | -0.13*** | -0.29*** | 0.10** | 0.16*** | 0.04 | 0.15** | |
| OSM | Nodes per Capita | Quantity | 0.07 | -0.12* | 0.17*** | 0.16*** | 0.02 | 0.19*** | |
| OSM | Ways per Capita | Quantity | 0.28*** | -0.04 | 0.14*** | 0.15*** | 0.03 | 0.31*** | |
| OSM | Tags (Nodes + Ways) per Capita | Quantity | 0.17*** | -0.09 | 0.17*** | 0.15*** | 0.03 | 0.26*** | |

**Table 1: The results of our univariate and multivariate regressions for Wikipedia and OSM in both the United States and China. Each cell value represents the corresponding normalized beta coefficient. *** is p<0.001, ** is p<0.01, and * is p <0.05.**

It is important to point out the quality deficiencies in rural content are experienced directly by enormous numbers of people (and indirectly experienced through peer-production-based intelligent systems). According to the Wikimedia Foundation's page view statistics (collected via WikiBrain), every month, millions of people visit Wikipedia articles about places in very rural, U.S. counties (and this does not include the people who look at OSM-based maps about these places). Indeed, we aggregated all pages views to all articles about places in each county over a one-month period, and found that the median, very rural county (NCHS classification = "6") received over 6923 page views (NCHS = "5" counties had a median of 15600). These figures are not a surprise: in the United States, over 46.2 million people live in rural areas [77], and many others need information about these areas.

**Theme 2: Rural Advantage in Per Capita Quantity**
Whereas nearly all of our quality results point to a strong urban advantage, Table 1 shows that the opposite is true of the *quantity* of this content. Nearly all of the quantity attributes (Type = "Quantity") in Table 1 show a strong and significant negative effect for *% Pop Urban*, indicating a substantial rural advantage in the per capita quantity of peer-produced information.

Examining the *Articles per Capita*, *Nodes per Capita*, and *Ways per Capita* figures, we see that there are indeed many more features of interest in rural areas than in urban areas. For instance, in core urban counties, there is an average of 2,238 potential local editors per article, whereas in purely rural counties this number drops to 467. Given that a miniscule percentage of the populace edits Wikipedia or contributes to OSM [23], this places a tremendous burden on local contributors in purely rural counties.

Much of the past work (e.g., [4,19,27,38]) that examines content biases implicitly or explicitly assumes that the distribution of content should be roughly equal on a per capita basis, e.g., that a city of, say, 100,000 in sub-Saharan Africa should be described by roughly the same number of articles with roughly the same total length as a city of 100,000 in California. Another important trend in Table 1 problematizes this assumption. For instance, consider the United States *Length (Bytes) per Capita* and *Tags (Nodes + Ways) per Capita* rows. Under the "equal per capita" assumption, we would assume that Wikipedia is tremendously biased towards rural areas, and perhaps that dramatic effort is needed by the community to reduce this bias. However, considering these quantity results in context of the quality results, we see that much of this content is generated by bots, batch editors, and fly-by editors who do not focus in the local area and contribute far less rich local content. Indeed, the end result is that there is actually less content from locally focused editors and the overall proportion of quality articles is less, even on a per capita basis. While raw content is certainly important from some perspectives, this work suggests that it may not tell the whole story.

It is important to note that there is a significant outlier to the general rural advantage in quantity: OSM in China. *In China, there are more ways and tags per capita in urban areas than in rural areas*. A quick examination of the raw data revealed the cause of this outlier: the OSM tools that import massive amounts of spatial data about rural areas in the United States cannot function in China due to government restrictions on the datasets required by these tools. Chinese state law stipulates that geographic data of a certain level of accuracy and scale should be kept secret for national security reasons [48]. As such, OSM in China provides a window into urban/rural peer production dynamics when bots and batch editing (e.g., that help to upload government data) do not exist. Without bots and batch editors, not only is the quality of peer-produced content worse in rural areas, but also, in many cases, this content simply does not exist. We return to this issue in the discussion section.

**Theme 3: Trends in the Control Variables**
In addition to revealing strong, significant, and persistent effects for urban/rural dynamics, our modeling efforts also unexpectedly revealed similar effects for some control variables. Most notably, in the United States, Table 1 shows that more Democratic counties and counties with a higher *% White Collar* tend to have higher-quality content. For instance, unpacking the effect sizes in Table 1, we see that for every standard deviation (14.7%) increase in the Obama 2012 vote rate, there is a corresponding 20.6% increase in the percentage of articles that are C-class or better (multivariate model). The control variable results in Table 1 provide important motivation for future work that can focus on the corresponding phenomena in more detail.

## DISCUSSION AND IMPLICATIONS
The results in this paper have important implications for peer production communities' efforts to describe rural phenomena, as well as the peer production content generation model more generally.

**Systemic Challenges in Describing Rural Areas**
Our results point to two distinct challenges for efforts to generate high-quality content about rural phenomena in peer production communities. The first is that rural participation in these communities seems to be lower than urban participation, at least when it comes to contributing content about their home areas. In particular, we observed that there are fewer tokens from locally-focused editors *per capita* in articles about rural areas, with these tokens much more likely to be rich local information than tokens from non-locally focused editors. Reduced participation and its corresponding effects on quality are likely one reason why rural areas have fewer C-class and above articles on a per capita basis than urban areas.

Uneven participation in peer production communities and the corresponding deficiencies in content have been observed in a number of domains, in particular gender. By adapting some of the techniques that have been used to bring women and other underrepresented groups into peer production editor communities (e.g., [68]), this rural participation problem could possibly be partially mitigated.

The second systemic challenge facing peer production communities in describing rural phenomena, however, is much harder to address. Put simply, our results suggest that (1) *it is far more difficult to describe many rural phenomena using peer production than it is urban phenomena* and (2) *the increased difficulty systematically leads to lower quality peer-produced content about rural areas*.

Let us examine this challenge in more detail. Consider, for example, the task of generating a quality article about every incorporated place (e.g., city, town) in the United States. This is a task that Wikipedia has taken on, as the community has determined that all incorporated places are sufficiently notable so as to deserve an article. We know from prior work that a large percentage of contributions to peer production repositories come from locals [26,29], and we also know from prior work [8,72] and the coding study above that local contributions tend to be of higher quality than those of non-locals. As such, for some incorporated places – e.g., New York City – there are literally millions of potential local experts who can help create a high-quality article about the incorporated place. However, for other incorporated places – e.g., Orrtanna, PA (pop. 173) – this number is much smaller. Indeed, we saw above that while there are over 2,200 potential local editors for every Wikipedia article about core urban areas, there are less than 500 in very rural areas.

Given this systemic challenge, it is not a surprise that we found that peer-produced content about rural areas tends to be of much lower quality. Stated more formally, our results suggest the following general property of current models of geographic peer production:

> *Peer-produced data about rural areas will be of lower quality when the ratio of entities of interest to the size of the local population is much higher in rural areas than in urban areas.*

In addition to incorporated places, countless geographic phenomena have more entities-per-capita in rural areas than in cities, and many of these phenomena are being mapped/described in Wikipedia and OSM: e.g., roads, counties, schools, natural phenomena (e.g., creeks, rivers). While not all geographic phenomena display this property – U.S. congressional districts, for instance, are population-controlled – many do. It is reasonable to expect that if rural areas can increase participation rates, they may be able to compete with urban areas on phenomena like congressional districts. However, rural participation rates would have to be orders of magnitude higher than that of urban areas to generate widespread quality content about phenomena for which the number of entities of interest per capita is substantially higher in rural areas (e.g., as with incorporated places).

It is important to note that the basic principle behind the general property of rural peer production stated above is not a new idea. Indeed, in some ways, the general property is an instance of Linus' Law ("Given enough eyeballs, all bugs are shallow" [54]), which has been explored and characterized in peer production in other contexts (e.g., [9,22]). However, our results highlight the fact that, due to the inherent, low-population-density nature of rural areas, peer production communities will in general find themselves on the wrong end of Linus' Law when trying to describe or map rural phenomena.

**Different Model of Peer Production for Rural Areas**

Automated and partially-automated software agents that generate content can be quite controversial in peer production communities [40,71]. However, our China OpenStreetMap results show what happens to rural peer-produced content without these agents: not only is content of lower quality, but also, in many cases, it simply does not exist. While relying solely on high-quality manual edits may be possible for content about urban areas, our research demonstrates that this is not true for rural areas.

More generally, our results point to descriptions of rural areas benefiting from a different model of peer production than exists in cities. In this model, bots and batch editing play a more central role to partially account for the reduced amount of local expertise per entity of interest. By embracing this peer production model and developing new technologies to support it, it may be possible to increase the quality of content about rural areas, especially as automated content generation technologies become more effective.

For example, new tools (e.g., Reasonator [42]) are becoming available that attempt to turn information from Wikidata, Wikipedia's structured data sister project, into natural language Wikipedia articles. Wikidata supports information well beyond that available in government sources like a census, allowing Reasonator tools to generate text that more resembles that contributed by editors with a local focus (e.g., information about a town's mayor or its prominent citizens). The Wikipedia community has been heavily critical of incorporating content from Reasonator and related technologies, but content about rural areas may benefit significantly from this content in the future. Additionally, because quality in rural areas is already low, rural articles provide a useful do-no-harm (or do-little-harm) testing ground for these technologies.

Similar approaches in OSM are also possible. For example, automated approaches could be developed to extract semantic information (e.g., opening hours) from business' websites and assign this information as tags to the corresponding OSM entity. Computer vision operating on satellite imagery may also help increase rural data quality.

**LIMITATIONS AND FUTURE WORK**

The research presented above presents a number of opportunities for follow-on work:

(1) While this research examined an Eastern country and a Western country, our results should be confirmed in a variety of other cultural contexts as well. Moreover, a more qualitative investigation of the complex rural/urban editing choices being made in each online community and in each cultural context would be quite informative.

(2) It may be useful to consider our findings in the context of long-standing discussions about whether a particular class of entity is sufficiently notable for Wikipedia articles (e.g., [76]). When the entity class under consideration is geographic in nature, tools based on our work can inform this debate by predicting expected quality in urban and rural areas.

(3) An exciting area of future work arises out of the possibility of encouraging urban contributors to "adopt" a rural region, learn about that region, and become local specialist editors in that region, even if they do not live there (although the effectiveness of this approach would have to be measured carefully).

**CONCLUSION**

Examining both Wikipedia and OpenStreetMap in both China and the United States, this research showed that peer production faces systemic challenges in describing rural phenomena, challenges that will persist even if the observed participation issues are addressed. More generally, this work adds to a growing body of literature that suggests that urban/rural dynamics play a key role in geographically-referenced content that is produced in online social systems. We hope our results encourage further investigation of these dynamics, as well as the development of tools and strategies to help mitigate the identified problems.

**ACKNOWLEDGEMENTS**

We would like to thank Shilad Sen for his contributions to the WikiBrain project that greatly assisted with this work. We would also like to thank Morten Warncke-Wang, Jacob Thebault-Spieker, Subhasree Sengupta, Taylor Long, Sarah McRoberts, Joe Konstan, Loren Terveen, and the rest of our GroupLens colleagues for brainstorming with us and helping to facilitate data collection. This project was supported by NSF IIS-1526988, NSF IIS-1421655, a Wikimedia Foundation Engagement Grant, a Google Research Faculty Award, FWO K207615N, and a University of Minnesota College of Science and Engineering Graduate Fellowship.